\documentclass[12pt,preprint]{aastex}

\shorttitle{Very Hot Jupiters Thermal Emission }
\shortauthors{L\'opez-Morales \& Seager}

\begin{document}

\title{Thermal Emission from Transiting Very-Hot Jupiters: Prospects for Ground-based Detection at Optical Wavelengths}

\author{Mercedes L\'opez-Morales\altaffilmark{1} \& Sara Seager\altaffilmark{2}}
\email{mercedes@dtm.ciw.edu, seager@mit.edu}

\altaffiltext{1}{Carnegie Fellow. Carnegie Institution of Washington, Department of Terrestrial Magnetism, 5241 Broad Branch Rd. NW, Washington D.C., 20015, USA}
\altaffiltext{2}{Department of Earth, Atmospheric, and Planetary Sciences, Department of Physics, Massachusetts Institute of Technology, 77 Massachusetts Ave., Cambridge, MA 02139-4307 USA}

\begin{abstract}
Very hot Jupiters (VHJs) are defined as Jupiter-mass extrasolar
planets with orbital periods shorter than three days. For low albedos
the effective temperatures of irradiated
VHJs can reach 2500--3000 K. Thermal emission from VHJs is therefore
potentially strong at optical wavelengths.  We explore the prospects
of detecting optical-wavelength thermal emission during secondary
eclipse with existing ground-based telescopes.  We show
that OGLE-TR-56b and OGLE-TR-132b are the best suited candidates for
detection, and that the prospects are highest around ${\it z'}$-band
($\sim$ 0.9 $\mu$m). We also speculate that any newly discovered VHJs
with the right combination of orbital separation and host star
parameters could be thermally detected in the optical. The lack of
detections would still provide constraints on the planetary albedos and
re-radiation factors.

\end{abstract}


\section{Introduction} \label{sec:intro}

Five hot Jupiters have published lower atmosphere measurements
(Charbonneau et al. 2002, 2005; Deming et al. 2005, 2006; Harrington
et al. 2006; Knutson et al. 2007; Grillmair et al. 2007; Richardson et
al.  2007; Tinetti et al. 2007).  Several upper atmosphere detections
have also been reported (Vidal-Madjar 2003, 2004; Ballester et
al. 2007).  All these measurements have been made from space, with the
{\it Spitzer} and {\it Hubble} telescopes.  Exoplanet atmosphere
detection from the ground has so far been elusive, despite many tries
(e.g. Deming et al. 2007; Snellen \& Covino 2007; Snellen 2005). Yet,
ground-based detection capability would greatly facilitate studies of
hot Jupiter atmospheres.

Hot Jupiters are Jupiter-mass planets in 3 to 9 day period
orbits. Very hot Jupiters (VHJs) are those in 1 to 3 day period
orbits. These planets are heated primarily from incident radiation
from their parent stars. The closer the planet is to the star, the
hotter the planet will be. VHJs orbiting solar-type stars may have
effective temperatures, $T_{p}$, as high as 2500--3000K. Therefore,
VHJs may be as hot as brown dwarfs or very low-mass stars. In this
case, the contribution of the thermal emission from the planet to the
total light of the planet-star system will be significantly high, and
could be detected in the optical or near-infrared during secondary
transits.

Eleven transiting VHJs are currently known (see Table 1). 
We explore the thermal radiation contribution of these 
planets at wavelengths between V and K-band (0.5--2.5$\mu$m), and the 
prospects of detecting their secondary transits  with current
ground-based instruments. Our study focuses on the thermal
emission from the planets' lower atmosphere, defined to be in the 
1 to $10^{-5}$ bar pressure range.

\section{Thermal Emission vs. Reflected Light} \label{sec:temps}

For VHJs at optical wavelengths one usually thinks of the reflected
light without considering that, depending on the VHJ effective
temperature, thermal emission could dominate reflected light.

The effective temperatures of externally heated extrasolar planets can be
estimated from energy balance as
\begin{equation}
T_{p} = T_{*} \left(\frac{R_{*}}{a}\right)^{1/2} [f(1-A_{\rm B})]^{1/4} ,
\end{equation}
where $T_{p}$ is the temperature of the planet, $T_{*}$ and
$R_{*}$ are the effective temperature and the radius of the star, $a$
is the star-planet orbital separation, and $f$ and $A_{\rm B}$ are
the re-radiation factor and the Bond albedo of the planet. 

Given the planets effective temperatures, their thermal and reflected 
fluxes can be computed using
\begin{equation}
F_{p_{th}} = \frac{2h\nu^{3}}{c^{2}} \frac{\pi R_{p}^{2}}{e^{\frac{h\nu}{kT_{p}}} - 1} \frac{1}{D^{2}},
\end{equation}
for the thermal flux, and
\begin{equation}
F_{p_{ref}} = F_{*} A_{g} \frac{R_{p}^{2}}{a^2} ,
\end{equation}
for the reflected flux, where $F_{*}$ is the stellar surface
flux divided by $D^{2}$, $A_{g} = \frac{2}{3} A_{B}$, assuming 
Lambert's law (e.g., Rowe et al. 2006), and $R_{p}$ is the radius of the planet.
$D$ is the distance from the planet to Earth.

\subsection{Effective Temperatures}

We computed temperatures for each known VHJ from eq. (1), adopting
published masses, radii, and effective temperatures for the host stars
(see Table~1).  The orbital separations were derived
from the mass of the stars and the planets, and the published orbital
periods.

The parameter $f$ describes how the stellar radiation absorbed by a
planet is redistributed in its atmosphere. The absorbed
radiation can be reradiated back to space, advected around the
planet, or a combination of both. If the radiative timescale is
shorter than the advective timescale, the incident radiation will be
reradiated back to space, in which case $f=2/3$. In the opposite case,
atmospheric circulation redistributes the energy around the planetary
atmosphere before it can get reradiated, and $f = 1/4$.  Recent
studies of the day-to-night side brightness variations of $\upsilon$
Andromedae b (Harrington et al. 2006), and HD189733b (Knutson et
al. 2007), indicate that both scenarios are possible. We
consider values of $f$ = 2/3 -- 1/4 to bracket the possible range.

The Bond albedo $A_{B}$ is the fraction of incident stellar radiation
reflected by the planets atmosphere. $A_{B}$ depends on a
planet's chemical composition, which in turn depends partly on
temperature.  We consider the cases $A_{B}$ = 0, 0.3, and
0.5. Observations indicate that the albedos of hot Jupiters are in
fact very low ($A_{B}$ $<$ 0.15; Rowe et al. 2006; Harrington,
priv. comm.). Model atmospheres also suggest these planets are dark
(e.g. Marley et al. 2007). Even so, we include $A_{B}$ = 0.3 and 0.5 
to illustrate
possible unexpectedly reflective cases.  VHJs may be as bright as
$A_{B}$ = 0.3 if covered by homogeneous pure silicate clouds at one
millibar level or higher altitudes (Marley et al. 1999, Seager et
al. 2000, Sudarsky et al. 2000).  If they have patchy silicate or iron
clouds, then $A_{B}$ will be $<$ 0.3, as the stellar radiation can
penetrate below the clouds and be absorbed by gas-phase molecules
(Hood et al. 2007; submitted). Of most relevance to this work is that VHJ
atmospheres are too hot on the substellar side for silicate or iron
clouds to form.

Table 1 gives the expected $T_{p}$ for each known transiting VHJ, for
the $f$ and $A_{B}$ values above. The table includes formal errors
derived from eq.~1 and published parameters of each system.  The
maximum expected temperatures in the table reveal two groups of
VHJs. One with $T_{p_{Max}}$ $>$ 2600K, that includes OGLE-TR-56b and
OGLE-TR-132b, and a second group with $T_{p_{Max}}$ $<$ 2200K, that
includes the other nine planets. OGLE-TR-56b and OGLE-TR-132b are the
most likely candidates for detection.

\subsection{Thermal Emission and Reflected Light Fluxes}

A successful detection of radiation from transiting extrasolar planets 
depends on the ratio of fluxes emitted by the planet and the star 
at a given wavelength. Figure~1 shows the computed model stellar fluxes for
OGLE-TR-56 and OGLE-TR-132, between 0.5 and 2.5$\mu$m, and their 
expected thermal emission and reflected light fluxes,
based on the $T_{p}$ and $A_{B}$ values from \S 2.1.

The stellar fluxes are derived using grids of Kurucz (1993) models,
interpolated for the specific $T_{eff}$, [Fe/H], and log$g$ values of
OGLE-TR-56 and OGLE-TR-132. Stellar parameters for OGLE-TR-56 have
been measured by Santos et al. (2006), who obtain $T_{eff}= 6119 \pm$
62K, [Fe/H] = +0.25 $\pm$ 0.08 dex, and log$g$ = 4.21 $\pm$ 0.19 cgs.
For OGLE-TR-132, the stellar parameters derived by Bouchy et
al. (2004) are $T_{eff}= 6411 \pm$ 179 K, [Fe/H] = +0.43 $\pm$ 0.18
dex, and log$g$ = 4.86 $\pm$ 0.14 cgs.

For the planets, we derive thermal emission and reflected light fluxes
from eqs. (2) and (3), assuming the planets and the stars emit as
blackbodies.  We consider two temperature scenarios: 1) $f$ = 1/4 and
$A_{B}$ = 0.5, where the planetary fluxes in the optical are dominated
by reflected light (such high albedos are unlikely for VHJs, see \S
2.1), and 2) $f$ = 2/3 and $A_{B}$ = 0.05, close to the $f$ = 2/3;
$A_{B}$ = 0.0 scenario, to illustrate the significantly lower
contribution of reflected light versus thermal emission for very low
albedos. The expected temperatures of OGLE-TR-56b and OGLE-TR-132b in
scenario 2) are 2852 $\pm$ 24K and 2581 $\pm$ 35K.

We go beyond blackbody models by including in Fig.~1 the Hubeny et al.
(2003) models for irradiated planets at $T_{p}$ = 2600K, with and
without TiO/VO molecules in their atmospheres. In the latter case, TiO
and VO have condensed into solids, no longer
contributing to the opacity. TiO and VO are such strong absorbers that
the incident stellar radiation is absorbed at an altitude where
reradiation dominates over advection, potentially leading to very hot
planetary atmospheres on the substellar side. The presence of TiO and
VO is furthermore expected to create a temperature inversion in the
lower atmosphere in some cases (e.g., metal rich, see also Fortney et
al. 2006), leading to emission lines and making the planet flux at
some wavelengths brighter than blackbodies of the same temperature.

\section{Detectability of Secondary Transits at Optical Wavelengths}\label{sec:anal}

Secondary transits can be detected at any given wavelength if the 
planet-to-star flux ratios, $F_{p}/F_{*}$, are high enough.
In terms of 
magnitudes, the transits are detectable if the difference in magnitude during
transit, $\mid\Delta$mag$\mid$, is larger than the photometric precision of
the observational light curves, $\sigma_{mag}$. $\mid \Delta$ mag $\mid$ is
derived as 
\begin{equation}
  \mid\Delta mag \mid = 2.5\cdot log_{10} \left(1 + \frac{F_{p}}{F_{*}}\right),
\end{equation}
where $F_{p}$ is either the thermal emission flux, $F_{p_{th}}$, or the 
reflected light flux, $F_{p_{ref}}$. Detectability can be further increased
by binning the transit data, as done for example by Deming et al. (2005) for
HD209458b. The photometric precision of the light curves is then
$\sigma_{mag}$/$\sqrt{t}$, where $t$ is the time duration of the bins.

Figure 2 shows the expected $\mid\Delta$mag$\mid$ during secondary
transits of OGLE-TR-56b and OGLE-TR-132b, for the thermal emission and
reflected light cases in Fig.~1. A detailed description is given in
the figure's caption.  The stellar and planetary fluxes have been
integrated over two model filter passbands with FWHMs = 0.1$\mu$m,
centered at 0.76$\mu$m and 0.91 $\mu$m. Those passbands resemble the
$i'$-- and $z'$--band filters of the Sloan Digital Sky Survey (SDSS;
Fukugita et al. 1996).

We also show the achievable $\sigma_{mag}$ for OGLE transit light
curves at optical wavelengths (horizontal dashed lines in Fig.~2)
for integration times of $t$ = 1, 120 (3$\sigma$),
and 120(1$\sigma$) minutes. The transits of both planets last about
120 min, so the two bottom lines represent the 1$\sigma$ and 3$\sigma$
level photometric precision that would be achieved by binning
$\sigma_{mag}$ = 1.0 mmags/min light curve data over the entire
duration of the transits (see below and  \S 4). 

To compare our estimates with $\sigma_{mag}$ from OGLE primary
transit light curves, we give the average $\sigma_{mag}$ and exposure
times per data point in Table~2. Assuming no correlation between
photometric precision and filter passbands, $\sigma_{mag}$ varies
between 0.83 and 1.18 mmags/min, depending on the instrument and the
target.  Here we have excluded the Magellan/IMACS values because its
low photometric precision results from instrumental effects (Winn et
al. 2007b).  We therefore reasonably assume an average $\sigma_{mag}$
of 1.0 mmags/min.

\section{Discussion} \label{sec:disc}
Secondary transits of extrasolar planets are normally pursued in the infrared,
where $F_{p_{th}}/F_{*}$ is higher than in the optical. We argue that, for the 
right combination of stellar and planetary
parameters, the thermal emission from some VHJs could be detected at optical 
wavelengths during secondary transits.

Our investigation of all known transiting VHJs concludes that only two of
them, OGLE-TR-56b and OGLE-TR-132b, have the right combination of
stellar and planetary parameters to be hotter than 2600K.
At those temperatures thermal emission dominates over reflected light
emission.  Therefore thermal emission could be detected during
secondary transits in the optical. OGLE-TR-56b is the most promising
candidate for detection, in either $i'$-- or $z'$--band. OGLE-TR-132b
could also be detected in $z'$--band, but the prospects are
lower. Reflected light emission is similar in both filters, and
a factor of 10--20 less than the
thermal emission if the planets are very hot.

Observations in $z'$--band are better suited for
detecting thermal emission from these two planets than ground-based 
observations at other optical or near-IR passbands.
The better performance of $z'$--band versus $i'$--band is clearly illustrated 
in Fig.~2. At shorter wavelengths, e.g. 0.50--0.65$\mu$m, 
$F_{p}/F_{st}$ decreases considerably (see Fig.~1). At near-IR 
wavelengths, the expected $F_{p}/F_{st}$ are higher, but the effect
of the atmosphere on
ground-based observations produce lower quality light curves 
(Snellen \& Covino 2007; D\'iaz et al. 2007). 

The transits of OGLE-TR-56b and OGLE-TR-132b last for about two hours.
By binning $z'$-band light curves with $\sigma_{mag}$ 
= 1.0 mmag/min over the duration of the transits (in-transit and out-of-transit
 bins of the same duration), thermal emission from 
OGLE-TR-56b could be detected with $>$ 5$\sigma$ significance
after three secondaries, if the planet emits as a blackbody with
$T_{p}$ $>$ 2500K. If the planet emits as predicted by the Hubeny et al.
(2003) models with TiO/VO, the expected transit depth is 0.04\% 
 (filled magenta squares in Fig.~2). The transit could be then
detected with $>$ 6$\sigma$ significance after three transits. In the case
of OGLE-TR-132b, four secondary transits are needed for a 5$\sigma$ 
significance detection.

One might expect that cooler VHJs around brighter stars would be
as promising or more so than the more distant and fainter (yet
hotter) OGLE-TR-56b and OGLE-TR-132b.  In principle, slightly cooler
planets around brighter stars should be just as favorable, because the
brighter stars have lower Poisson photon noise.  In practice,
$\sigma_{mag}$ is limited for bright stars in two ways.  On small
telescopes atmospheric scintillation limits $\sigma_{mag}$ 
from reaching the photon noise limit ($\sigma_{mag}$
from scintillation  is typically several
millimags). On large telescopes the limit is usually imposed by the absence
of suitable nearby comparison stars within their small fields of view.

Any newly discovered transiting VHJs should be examined for the
possibility of detecting their thermal emission at optical
wavelengths. In particular hot VHJs orbiting bright stars with a
binary star companion would be ideal.  Such detections will provide
important clues about the energy processing mechanisms undergoing in
the atmospheres of those planets.

\acknowledgments
M.~L-M. acknowledges support from the Carnegie Institution of 
Washington through a Carnegie Fellowship, and from the NASA Astrobiology 
Institute. We thank A. Bonanos for her help generating 
Kurucz models and I. Hubeny for helpful clarifications on his models.
We also thank an anonymous referee for helpful suggestions.

\clearpage

\begin{table}
\caption{Effective temperatures of the eleven known transiting VHJs, for $A_{B}$ = 0.0, 0.3, and 0.5, and $f$ = 2/3 and 1/4.}
\label{tab:AbsDim} 
\scriptsize{
\begin{tabular}{rccc|cccr}
\hline\hline
Planet& $T_{p}(K)$ & $T_{p}(K)$ & $T_{p}(K)$ & $T_{p}(K)$ & $T_{p}(K)$ & $T_{p}(K)$ & Ref.\tablenotemark{1}\\
      & f=2/3; $A_{B}=0$ & f=2/3;$ A_{B}=0.3$ & f=2/3; $A_{B}=0.5$ & f=1/4; $A_{B}=0$ & f=1/4; $A_{B}=0.3$ &f=1/4; $A_{B}=0.5$ \\

\hline

OGLE-TR-56b  & 2889 $\pm$ 24& 2642 $\pm$ 22& 2429 $\pm$ 20 & 2260 $\pm$ 19& 2068 $\pm$ 17& 1901 $\pm$ 16& 1,2\\
OGLE-TR-113b & 1717 $\pm$ 14& 1570 $\pm$ 13& 1444 $\pm$ 12 & 1344 $\pm$ 11& 1229 $\pm$ 10& 1130 $\pm$ 9 & 1,3,4\\
OGLE-TR-132b & 2615 $\pm$ 36& 2392 $\pm$ 32& 2199 $\pm$ 30 & 2046 $\pm$ 28& 1872 $\pm$ 25& 1721 $\pm$ 23& 5,6\\
HD189733b    & 1500 $\pm$ 10& 1372 $\pm$ 9 & 1261 $\pm$ 8  & 1174 $\pm$ 8 & 1074 $\pm$ 7 &  987 $\pm$ 6 & 7,8\\
XO-2b	     & 1682 $\pm$ 15& 1539 $\pm$ 14& 1415 $\pm$ 13 & 1316 $\pm$ 12& 1204 $\pm$ 11& 1107 $\pm$ 10& 9\\
Corot-exo-1b & 2225         & 2036         & 1871          & 1742         & 1593         & 1464         & 10\\
WASP-1	     & 2177 $\pm$ 61& 1991 $\pm$ 56& 1831 $\pm$ 51 & 1704 $\pm$ 48& 1558 $\pm$ 44& 1433 $\pm$ 40& 11\\
WASP-2	     & 1615 $\pm$ 96& 1478 $\pm$ 87& 1358 $\pm$ 80 & 1264 $\pm$ 75& 1156 $\pm$ 68& 1063 $\pm$ 63& 11\\
HD140926b    & 2226 $\pm$ 30& 2036 $\pm$ 27& 1872 $\pm$ 25 & 1742 $\pm$ 23& 1593 $\pm$ 21& 1465 $\pm$ 20& 12\\
TrES-2	     & 1882 $\pm$ 15& 1722 $\pm$ 14& 1583 $\pm$ 13 & 1473 $\pm$ 12& 1347 $\pm$ 11& 1238 $\pm$ 10& 13\\
TrES-3	     & 2100 $\pm$ 32& 1921 $\pm$ 29& 1766 $\pm$ 27 & 1643 $\pm$ 25& 1503 $\pm$ 23& 1382 $\pm$ 21& 14\\

\hline\hline
\end{tabular}
}
\tablenotetext{1}{[1] Santos et al. (2006), [2] Pont et al. (2007), [3] Gillon et al. (2006),
 [4] D\'iaz et al. (2007), [5] Bouchy et al. (2004), [6] Moutou et al. (2004), [7] Bouchy et al. (2005),
 [8] Winn et al. (2007a), [9] Burke et al. (2007), [10] European Space Agency (ESA) press release, [11] Charbonneau et al. 2007, [12] Sato et al. (2005),
[13] Sozzetti et al. (2007), O'Donovan et al. (2007).}
\end{table}

\clearpage

\begin{table}
\caption{Parameters of the best published optical light curves of OGLE transiting planets.}
\footnotesize{
\begin{tabular}{rccccl}
\hline\hline
Planet & Central $\lambda$  & Telescope/Instr.  &Light Curve Dispersion  & Exposure Time &\\
       &      (\AA)       &                   &   (mmags)     &    (sec)   & Ref. Source\\
\hline
OGLE-TR-132b&6550&ESO VLT/FORS2\footnote{European Souther Observatory Very Large Telescope/FOcal Reducer and low dispersion Spectrograph 2}     &1.2      &55--60 &Moutou et al. (2004)\\
OGLE-TR-10b &8500&Magellan/MagIC\footnote{The Raymond and Beverly Sackler Magellan Instant Camera}&0.77     &55--85 &Holman et al. (2007)\\
OGLE-TR-111b&8100&Magellan/IMACS\footnote{Inamori magellan Areal Camera and Spectrograph}&1.5--2.0 &75--125&Winn et al. (2007b)\\
OGLE-TR-113b&6400&ESO NTT/SUSI2\footnote{New Technology Telescope/Superb Seeing Imager 2} &1.2--1.26&55     &Gillon et al. (2006)\\
OGLE-TR-56b &6550&ESO VLT/FORS1\footnote{European Souther Observatory Very Large Telescope/FOcal Reducer and low dispersion Spectrograph 1.}&0.9      &60\footnote{Time estimated from the 7-8 image cycles every 10 min in Pont et al. (2007), and assuming 22 sec readouts (Moutou et al. 2004)}     &Pont et al. (2007)\\
\hline\hline
\end{tabular}
}
\end{table}

\clearpage


\begin{figure}
\plotone{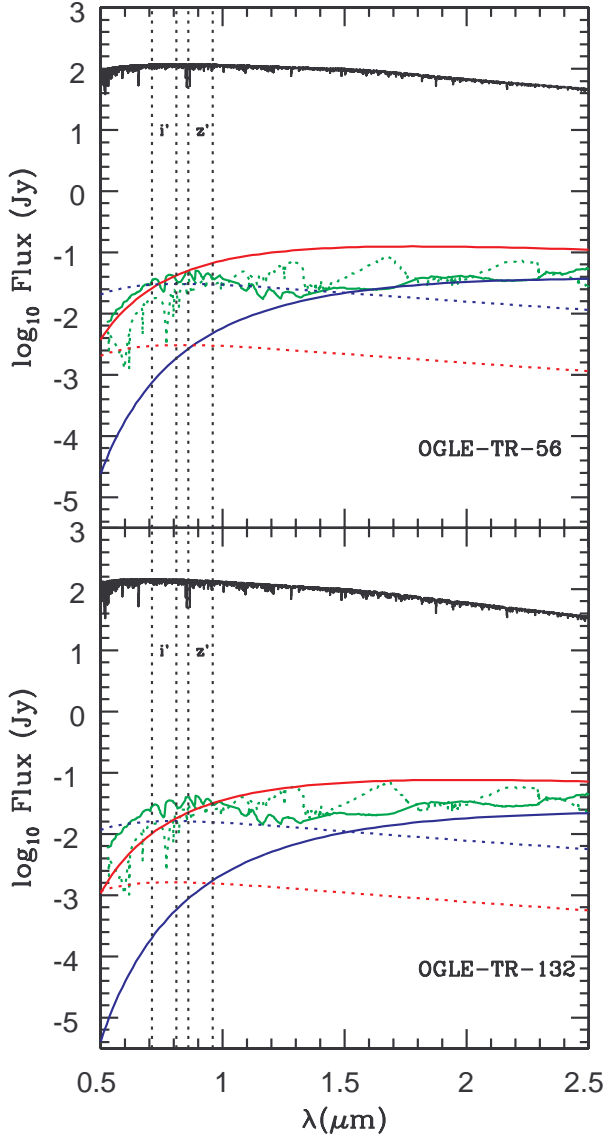}
\caption{Model stellar fluxes for OGLE-TR-56A-b (top) and OGLE-TR-132A-b (bottom) 
between 0.5 and 2.5$\mu$m. Black lines represent Kurucz (1993) models for the 
stars, 
multiplied by 4$\pi$ and $R_{st}^{2}/D^2$. Green lines show the
$T_{p}$= 2600K Hubeny et al. (2003) models with (solid line) and without 
(dotted line) TiO/VO. These models are multiplied by 4$\pi$ and 
$R_{p}^{2}/D^2$. The red and blue solid lines 
are the planets' blackbody thermal emission for $f$ = 2/3; $A_{B}$ = 0.05, and 
$f$ = 1/4; $A_{B}$ = 0.5 (see \S 2.2). The red and 
blue dotted lines are reflected light emission for the same two cases. 
All the fluxes are normalized to 
a distance $D=$ 10pc. $D$ is only a normalization factor and has no 
effect on the results in Fig. 2. For very low albedos $A_{B}$ and
inefficient energy redistribution factors $f$, the thermal emission 
of the planets dominates over their reflected light emission. The two vertical columns (dotted lines) show the $i'$ and $z'$ model passbands described in \S 3.}
\label{fig:lcs}
\end{figure}

\clearpage

\begin{figure}
\plotone{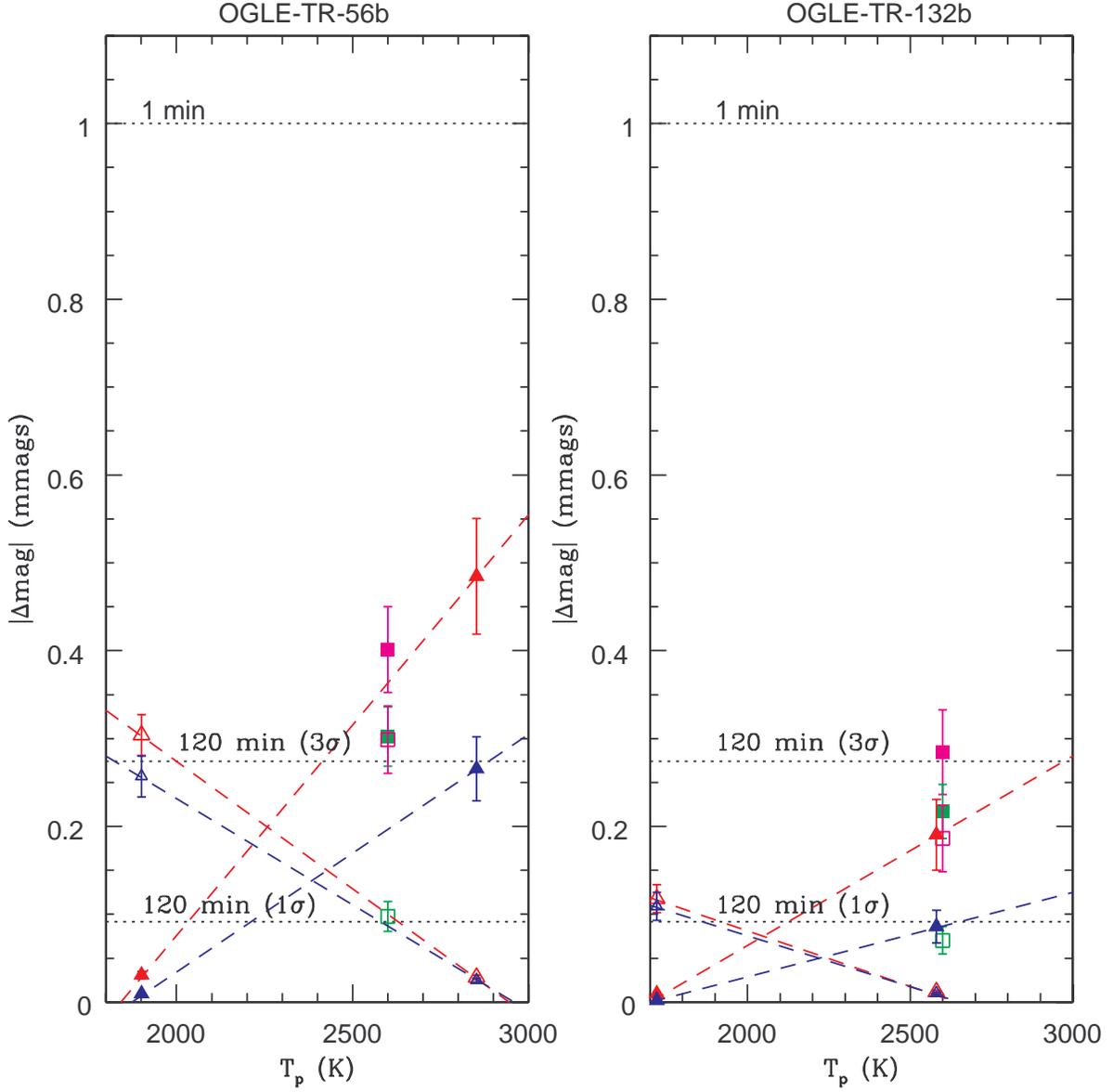}
\caption{Expected depths of secondary transits, $\mid\Delta$mag$\mid$, as a 
function of $T_{p}$ for OGLE-TR-56b (left-side) 
and OGLE-TR-132b (right-side). Filled triangles show the expected 
$\mid\Delta$mag$\mid$ if the planets emit thermally as blackbodies 
(see Fig.~1). 
Open triangles correspond to the cases of reflected light.
The dashed lines connecting the points guide the eye on how the value 
of $\mid\Delta$mag$\mid$ varies with the temperature of the planet $T_{p}$ 
in each case. Red illustrates
the results in $z'$--band; blue correspond to $i'$--band.
Green and magenta squares indicate the values of $\mid\Delta$mag$\mid$ if the 
planets emit as predicted by the Hubeny et al. (2003) models. Green shows the 
results in $i'$--band and magenta in $z'$--band. 
Filled and open symbols in this case show the results the models 
with and without TiO/VO. The errorbars on each point have been estimated by 
formal propagation of errors. Finally, the horizontal 
dotted lines indicate the photometric
precision achievable with current instruments on ground-based telescopes in 
t = 1, 120(3$\sigma$), and 120(1$\sigma$) minutes.}
\label{fig:1Ms}
\end{figure}

\end{document}